\documentclass[submission,copyright,creativecommons]{eptcs}
\providecommand{\event}{FROM 2022} 

\usepackage{iftex}

\ifpdf
  \usepackage{underscore}         
  \usepackage[T1]{fontenc}        
\else
  \usepackage{breakurl}           
\fi

\usepackage{amsthm}
\newtheorem{definition}{Definition}
\newtheorem{theorem}{Theorem}

\usepackage[mathscr]{euscript}

\usepackage{relsize}
\usepackage{amsmath}
\usepackage{amssymb}
\usepackage{oz}
\usepackage{csp-oz}
\usepackage{delarray}
\usepackage{slashed}
\usepackage{etex}
\usepackage{pgf}
\usepackage{tikz}
\usepackage{dsfont} 
\usepackage{float} 
\usepackage[noend,linesnumbered,ruled,vlined]{algorithm2e}
\usetikzlibrary{fadings}
\usepackage{setspace}
\usetikzlibrary{arrows,automata,shapes,patterns}
\usepackage{pgfplots}
\usepackage{filecontents}
\usepackage{pifont}
\usepackage{stackengine}
\usetikzlibrary{matrix}
\usepackage{multirow}

\SetAlFnt{\footnotesize}
\SetAlCapFnt{\footnotesize}
\SetAlCapNameFnt{\footnotesize}

\newcommand{\coop}[2][]{\langle\!\langle{#2}\rangle\!\rangle_{_{\!\mathit{#1}}}\,}

\SetKwFor{For}{for}{do}{EndLoop}
\SetKwFor{ForAllP}{loop forever do}{}{end}

\tikzset{
  treenode/.style = {align=center, inner sep=0pt, text centered,
    font=\sffamily},
  arn_o/.style = {thick,minimum height=1.3em,treenode, rectangle font=\sffamily\bfseries, draw=black,
    fill=orange, text width=1.3em},
     arn_bo/.style = {thick,minimum height=2.5em,treenode, rectangle font=\sffamily\bfseries, draw=black,
    fill=orange, text width=2.5em},
  arn_g/.style = {thick,minimum height=1.3em,treenode, rectangle, fill=mygreen, draw=black, 
    text width=1.3em },
      arn_so/.style = {thick,minimum height=0.75em,treenode, rectangle font=\sffamily\bfseries, draw=black,
    fill=orange, text width=0.75em},
  arn_sg/.style = {thick,minimum height=0.75em,treenode, rectangle, fill=mygreen, draw=black, 
    text width=0.75em },
  arn_xg/.style = {fill=mygreen, draw=black,thick,treenode, rectangle, draw=black,
    minimum width=0.75em, minimum height=0.75em},
     arn_xo/.style = {fill=orange, draw=black,thick,treenode, rectangle, draw=black,
    minimum width=1em, minimum height=1em},
}

\usepackage{url}
\urldef{\mails}\path|ntimm@cs.up.ac.za|

\newcommand{\BIGOP}[1]
{
\mathop{\mathchoice%
{\raise-0.22em\hbox{\huge $#1$}}%
{\raise-0.05em\hbox{\Large $#1$}}{\hbox{\large $#1$}}{#1}}}

\newcommand{\BIGboxplus}{\mathop{\mathchoice%
{\raise-0.35em\hbox{\huge $\boxplus$}}%
{\raise-0.15em\hbox{\Large $\boxplus$}}{\hbox{\large $\boxplus$}}{\boxplus}}}

\title{Synthesis of Cost-Optimal Multi-Agent Systems \\ for Resource Allocation}
\author{Nils Timm
\institute{University of Pretoria \\
 Pretoria, South Africa}
\email{nimm@cs.up.ac.za}
\and
Josua Botha
\institute{University of Pretoria \\
 Pretoria, South Africa}
\email{u19138182@tuks.co.za}
}
\def\titlerunning{Synthesis of Cost-Optimal Multi-Agent Systems for Resource Allocation}
\def\authorrunning{N. Timm \& J. Botha}
\begin{document}

\maketitle

\begin{abstract}
Multi-agent systems for resource allocation (MRAs) have been introduced  as a concept for modelling competitive resource allocation problems in distributed computing.  An MRA is composed of a set of agents and a set of resources.  Each agent has goals in terms of allocating certain resources.  For MRAs it is typically of importance that they are designed in a way such that there exists a strategy that guarantees that all agents will achieve their goals.  The corresponding model checking problem is to determine whether such a winning strategy exists or not,  and the synthesis problem is to actually build the strategy.  While winning strategies ensure that all goals will be achieved,  following such  strategies does not necessarily involve an optimal use of resources. 

In this paper, we present a technique that allows to synthesise cost-optimal solutions to distributed resource allocation problems. 
We consider a scenario where system components such as agents and resources involve costs.  A multi-agent system shall be designed that is cost-minimal but still capable of accomplishing a given set of goals.  Our approach synthesises a winning strategy that minimises the cumulative costs of the components that are required for achieving the goals.   
The technique is based on a propositional logic encoding and a reduction of the synthesis problem to the maximum satisfiability problem (Max-SAT). Hence, a Max-SAT solver can be used to perform the synthesis. From a truth assignment that maximises the number of satisfied clauses of the encoding a cost-optimal winning strategy as well as a cost-optimal system can be immediately derived. 
\end{abstract}

\section{Introduction}  
Multi-agent systems for resource allocation (MRAs) have been introduced in \cite{de2018generalising} as a concept for modelling competitive resource allocation problems in distributed computing.  An MRA is composed of a set of agents and a set of resources.  Each agent has a goals in terms of allocating certain resources for a time period before a deadline elapses.  Resources can be allocated by means of \emph{request} actions.  Further types of actions are \emph{release} and \emph{idle}.  MRAs run in discrete rounds.  In each round each agent selects an action, and the tuple of selected actions gets executed in a simultaneous manner.  Since resources are generally shared, the achievement of goals is a competition between agents.  
For MRAs (or more specifically, for the scenarios that they model) it is typically of importance that they are designed in a way such that there exists a strategy that guarantees that all agents will achieve their goals. 
In this context, a strategy is a mapping between states of the underlying MRA and actions to be taken by the agents in these states. 
The corresponding model checking problem is to determine whether such a winning strategy exists or not,  and the synthesis problem is to actually build the strategy. 
In \cite{timm2021model} we introduced a SAT-based technique for checking goal-achievability properties of multi-agent systems for resource allocation.  The technique does not only decide the model checking problem, it also synthesises a corresponding winning strategy if existent.  The approach encodes the  problem in propositional logic. Thus, model checking can be performed via satisfiability solving.  From a satisfying truth assignment of the encoded problem a winning strategy can be immediately derived.  


While winning strategies ensure that all goals will be achieved,  following such  strategies does not necessarily involve an optimal use of resources. 
In this paper, we extend our technique such that the outcome of the synthesis is not only a winning strategy but also a cost-optimal MRA solution of a distributed resource allocation problem.  We consider a scenario where a distributed system shall be designed that processes a given set of computational tasks.  Such a system is a composition of networked computers and shared resources.  System components need to be purchased.  Hence,  it is of interest to determine a cost-minimal system that is still capable of accomplishing all computational tasks. The distributed system to be designed can be straightforwardly modelled as an MRA where agents represent computers and the tasks are defined as goals of the agents. 
Given a set of tasks resp.  goals, we build an initial MRA that consists of sufficiently many agents and resources to achieve all goals.  In our new approach, prices can be assigned to agents and to the different types of resources in the system. 
The extended synthesis will build a \emph{cost-optimal} winning strategy:
Following such a strategy will ensure that all goals will be achieved, and it will additionally minimise the cumulative costs of the agents and resources that are actually \emph{used} in the run of MRA.  Hence, from a cost-optimal winning strategy we can derive a cost-optimal MRA by removing unused agents and resources. 
The cost-optimal MRA then indicates which components actually need to be purchased for the distributed system to be designed. 


The synthesis of cost-optimal strategies and systems is no longer a pure decision problem and therefore it cannot be reduced to standard Boolean satisfiability. 
However, we show that the synthesis can be reduced to the maximum satisfiability problem (Max-SAT).  Our reduction is based on an extension of the existing propositional logic encoding by weighted `\emph{not in use}' clauses. 
Each such clause encodes that a particular system component (agent or resource) is never used, and the weight of the clause corresponds to the price of the component. 
We have proven that from a truth assignment that maximises the sum of weights of satisfied `\emph{not in use}' clauses a corresponding cost-optimal winning strategy can be immediately derived. This allows us to employ Max-SAT solving  for synthesising cost-optimal strategies and MRAs. 

We first present our technique based on the case where goals are already assigned to agents and only the costs of the resources in the system need to be minimised. Secondly, we consider the case where goals are initially unassigned and both the costs of agents and resources need to be minimised.  We have implemented our approach on top of the Max-SAT solver OPEN-WBO \cite{martins2021open}.  
Experiments show promising results.
We demonstrate that optimal winning strategies synthesised via Max-SAT can involve significant cost savings in comparison to bare winning strategies synthesised via standard SAT. 

\section{Multi-Agent Systems for Resource Allocation }
In our approach we focus on synthesising strategies for  \emph{multi-agent systems for resource allocation }(MRAs), originally introduced in \cite{de2018generalising}.

\begin{definition}[Multi-Agent System for Resource Allocation]
A multi-agent system for resource allocation is a tuple $M = (Agt, T,Res,\$, \Phi)$ where
\begin{itemize}
 \item $Agt = \{a_1, \ldots, a_n \}$ is a   set of agents,
 \item $T = \{\tau_1, \ldots, \tau_m\}$ is a  set of resource types,
    \item $Res = \bigcup_{\tau \in T} Res_{\tau}$ is a   set of resources partitioned into subsets $Res_{\tau} = \{r_{\tau}^1, \ldots, r_{\tau}^{m_{\tau}}\}$ of resources of type $\tau$,
     \item $\$ : T \rightarrow \mathbb{N}$ is a price function that assigns a price to each type of resource; the price function extends to resources such that $\$(r) = \$(\tau)$ iff $r \in Res_{\tau}$,
\item $\Phi = \bigcup_{a \in Agt}\{\phi_{a}^1,\ldots,\phi_{a}^{g_a}\}$ is a   set of goals partitioned into subsets $\{\phi_{a}^1,\ldots,\phi_{a}^{g_a}\}$ of goals of agent $a \in Agt$, 
    \item each goal is a tuple $\phi_{a} = (R, p,d)$ where $R \subseteq T$ is the resource composition, $p \in \mathbb{N}$ is the period, and $d \in \mathbb{N}$ is the deadline of the goal.
\end{itemize}    
\end{definition} 
\noindent
Each agent has the objective to achieve all its goals.
A goal $\phi_{a} = (R,p,d)$ has been achieved 
if a resource of each type in $R$ has been allocated by agent $a$
for $p$ consecutive time steps before the deadline time step $d$ is reached. 

  \medskip
  \noindent
 \emph{Example. } We will illustrate the synthesis of strategies for achieving goal properties 
 based on the example system $M_{ex} = (Agt, T,Res,\$, \Phi)$ where
    \begin{itemize}
   \item $Agt = \{a_1, a_2, a_3\}$,
    \item $T = \{\tau_1, \tau_2, \tau_3\}$,
      \item $Res  = Res_{\tau_1} \cup Res_{\tau_2} \cup Res_{\tau_3}  = \{r_1, r_2 \} \cup \{r_3, r_4 \} \cup \{r_5, r_6 \}$,
      \item $\$(\tau_1) = 1$,  $\$(\tau_2) = 2$,  $\$(\tau_3) = 3$, 
      \item $\phi_{a_1}^1 = (\{\tau_1, \tau_2\},0,4)$,  $\phi_{a_1}^2 = (\{\tau_3\},0,1)$,
          \item $\phi_{a_2}^1 = (\{\tau_1, \tau_3\},0,4)$,
              \item $\phi_{a_3}^1 = (\{\tau_2\},0,1)$.
   \end{itemize}
   $M_{ex}$ consists of three agents and three different types of resources where two resources of each type are in place. 
The prices 1, 2 and 3 are associated with the different resource types. Agent $a_1$ has two goals whereas the agents $a_2$ and $a_3$ have one goal each.  The first goal of agent $a_1$ is to allocate one resource of type $\tau_1$ and one resource of type $\tau_2$ by time step 4. 
For simplicity, the periods of all goals are 0. That is, once all the required resources for a goal have been allocated they can be released in the next time step.

\medskip
\noindent
The actions that agents in MRAs can perform in order to achieve their goals are the following:
    
\begin{definition}[Actions]
Given an MRA $M$, the  {set of actions} $Act$ is 
the union of the following types of actions:
  \begin{itemize}
  \item {request actions}: $\{ req^a_r \mid a \in Agt,  r \in Res \}$
    
  \item  {release actions}: $\{rel^a_r  \mid a \in Agt,  r \in Res\}$

    \item  {release-all actions}: $\{ rel^a_{all} \mid a \in Agt\}$

  \item  {idle actions}: $\{ idle^{a} \mid a \in Agt\}$
  \end{itemize}
\end{definition}

\noindent
Hence, an agent can request a particular resource, release a particular resource that it currently holds, release all resources that it currently holds, or just idle.
An MRA runs in discrete rounds where in each round each agent chooses its next action.  In a round the tuple of chosen actions, one per agent, gets executed simultaneously.  The execution of actions leads to an evolution of the system between different states over time. 


\begin{definition}[States]
A state of an MRA $M$ is a function $s : Res \rightarrow Agt^+$ where $Agt^+ = Agt \cup \{a_0\}$ and $a_0$ is a dummy agent.
If $s(r) = a_0$ then resource $r$ is unallocated in state $s$. 
If $s(r) = a_i$ and $i > 0$ then $r$ is allocated by agent $a_i$ in $s$.
We denote by $s_0$ the initial state of $M$, where $s(r) = a_0$ for each $r \in Res$, i.e.  initially all resources are unallocated.
We denote by $S$ the set of all possible states of $M$.  If we want to express that resource $r$ is currently allocated by agent $a_i$ but the current state is not further specified, then we simply write $r = a_i$.
\end{definition}

\noindent
Hence, states describe the current allocation of resources by agents.
In each state only a subset of actions may be available for execution by an agent, which we call the protocol:

\begin{definition}[Action Availability Protocol]
  The {action availability protocol} is a function $P : S \times Agt \rightarrow 2^{Act}$ defined for each $s \in S$ and $a \in Agt$:
  \begin{itemize}
  \item   $req^a_{r}  \in P(s,a)$ iff $s(r)=a_0$;
  \item   $rel^a_{r}  \in P(s,a)$ iff $s(r)=a$;
     \item   $rel^a_{all} \in P(s,a)$ iff $ \vert s^{-1}(a)\vert > 0$;
    \item  $idle^a \in P(s,a)$. 
    \end{itemize}
  \end{definition}
  
  \noindent
Thus,  an agent can request a resource that is currently unallocated,  an agent can release one or all resources that it currently holds, and an agent can idle. 
  
  \begin{definition}[Action Profiles]
  An action profile in an MRA $M$ is a mapping $ap : Agt \rightarrow Act$.  $AP$ denotes the set of all action profiles. We say that a profile $ap$ is executable in a state $s \in S$ if for each $a \in Agt$ we have that $ap(a) \in P(s,a)$.
  \end{definition}
  
  \noindent
 Based on action profiles we can formally define the evolution of an MRA.
  
   \begin{definition}[Evolution]
The evolution of an MRA is a relation $\delta \subseteq S \times AP \times S$ where $(s, ap, s') \in \delta$ iff $ap$ is executable in $s$ and for each $r \in Res$:
\begin{enumerate}
\item if $s(r) = a_0$ then:
\begin{enumerate}
\item if $\exists  a : ap(a) = req^a_r \wedge \forall a' \neq a : ap(a') \neq req^{a'}_r$ then $s'(r) = a$;
\item otherwise $s'(r) = a_0$;
\end{enumerate}
\item if $s(r) = a$ for some $a \in Agt$ then:
\begin{enumerate}
\item if $ap(a) = rel^{a}_r \vee rel^{a}_{all}$ then $s'(r) = a_0$;
\item otherwise $s'(r) = a$.
\end{enumerate}
  \end{enumerate}
  \end{definition}
  
    \noindent
  If an action profile is executed in a state of an MRA $M$,  this leads to a transition of $M$ into a corresponding successor state, i.e.
    a change in the allocation of resources according to the actions chosen by the agents. According to the evolution, the request of a resource $r$ by an agent $a$ will be only successful if $a$ is the only agent that requests $r$ in the current round.
If multiple agents request the same resource at the same time, then none of the agents will obtain it. 

We are interested in solving strategic model checking problems with regard to MRAs: Do the agents in $Agt$ have a joint \emph{uniform strategy} that guarantees that all goals of all agents will be achieved?

         \begin{definition}[Uniform Strategy]
     A uniform strategy  of an agent $a \in Agt$ in an MRA is an injective function $\sigma_a : S \rightarrow Act$. 
A strategy can be also denoted by a relation $\sigma_a \subseteq S \times Act$ where $\sigma_a(s,act^a) = true$ iff $\sigma_a(s) = act^a$.
     A joint strategy for all agents in $Agt$ is a tuple of strategies $\sigma_{Agt} = (\sigma_{a_1},\ldots,\sigma_{a_r})$, one for each $a \in Agt$.
     We denote by $\Sigma_{Agt}$ the set of all possible joint strategies of $Agt$.
  \end{definition}

  \noindent
A strategy determines which action an agent will choose  in which state.  A strategy is uniform if the following holds: Each time when the system reaches the same state,  the agent will perform the same action according to the strategy.
  The outcome of a joint strategy $\sigma_{Agt}$ in a state $s$ is a path. 
  
   \begin{definition}[Outcome of a Strategy]
   Let $M$ be an MRA and $s$ a state of $M$.
   Moreover, let $\sigma_{Agt}$ be a joint strategy. Then the outcome of $\sigma_{Agt}$ in state $s$ is a path $\pi(s,\sigma_{Agt}) = s_0s_1\ldots$ where $s_0 = s$ and
\[
\bigwedge_{t \in \mathbb{N}}  \bigwedge_{a \in Agt} \Big( \sigma_a(s_t) \in P(s_t,a) \wedge \big(s_t,(\sigma_{a_1}(s_t) \ldots, \sigma_{a_n}(s_t)),s_{t+1}\big) \in \delta \Big).
\] 
We denote by $\Pi(M)$ the set of all possible paths of $M$.
   \end{definition}

   \noindent
   The MRA model checking problem is to decide whether a joint strategy exists that results in a path on which all goals will be achieved.  We call such a strategy a winning strategy.

        \begin{definition}[MRA Model Checking of Agent Goal Properties]
  Let $M = (Agt, T,Res,\$, \Phi)$ be an MRA, and let $s \in S$ be a state of $M$. 
 Then the strategic MRA model checking problem $[M,s \models \coop{Agt}\Phi]$ is inductively  defined
  as follows:        
  \[
  \begin{array}{lcllcl}
    
    [M,s \models \coop{Agt}\Phi] \ & \ \equiv \ & \ \bigvee_{\sigma_{Agt} \in \Sigma_{Agt}} \ [M,\pi(s, \sigma_{Agt}) \models  \Phi]    \vspace*{0.75mm} \ \\ 
    
    
        [M,\pi \models \Phi] \ & \ \equiv \ & \
\bigwedge_{\phi_a \in \Phi}    \       [M,\pi  \models  \phi_a]    \vspace*{0.75mm} \ \\ 


         [M,\pi \models \phi_a] \ & \ \equiv \ & \
\bigvee_{t=0}^{d-p} \bigwedge_{\tau \in R} \bigvee_{r \in Res_{\tau}} \bigwedge_{t'=t}^{t+p} \ \pi(t')(r) = a   
    \vspace*{0.75mm} \ \\ 
    
  \end{array}
  \]
  where $\phi_a$ is assumed to be the tuple $(R,p,d)$, 
and $\pi(t')$ denotes the $t'$-th state of the path $\pi$.
  \end{definition}
  
   \noindent
   Solving $[M,s \models \coop{Agt}\Phi]$ will not only decide the model checking problem but also synthesise a corresponding winning strategy $\sigma_{Agt}$ if existent.  
   
For our example system $M_{ex}$ the actions listed in Table 1 characterise a winning strategy.  Each action in the table is associated with an agent and a time step.  Agent $a_1$ performs action $req_{r_5}$ at time step 0, which immediately results in the achievement of the goal $\phi_{a_1}^2$.  Hence, the agent releases the allocated resource at the next time step. 
It then consecutively performs the actions $req_{r_1}$ and $req_{r_3}$ which results in the achievement of $\phi_{a_1}^1$.  The listed actions also ensure that all goals of the other agents will be achieved.

\begin{table}[h]
\caption{Actions that characterise a winning strategy.}
\begin{center}
\begin{tabular}{r|c|c|c|c|c}
\hline
 time step: & 0 & 1 & 2 & 3 & 4 \\ \hline
  $a_1$'s actions: & $req_{r_5}$ & $rel_{all}$ & $req_{r_1}$ & $req_{r_3}$ & $rel_{all}$ \\ \hline
  $a_2$'s actions: & $req_{r_2}$ & $req_{r_6}$ & $rel_{all}$ & $idle$ & $idle$ \\ \hline
   $a_3$'s actions: & $req_{r_3}$ & $rel_{all}$ & $idle$ & $idle$ & $idle$ \\ \hline
\end{tabular}
\end{center}
\end{table}

\noindent
   So far, costs of resources are not considered in the synthesis.

   \subsection{Resource Cost-Optimal Strategy Synthesis}
   We will now extend MRA model checking such that the outcome is a resource cost-optimal winning strategy.  For this, we first define the resource costs of paths.

  \begin{definition}[Resource Costs of Paths]
Let $M = (Agt, T,Res,\$, \Phi)$ be an MRA, 
and let $k = \textbf{\emph{max}}(d | (R,p,d)\in \Phi)$ be the latest deadline.
The cost of a path $\pi \in \Pi(M)$ with regard to a resource $r \in Res$ is defined as 
\[
c_r(\pi) \ = \ \left\{ \begin{array}{ll} 
     0 & \ \hbox{ if } \ \bigwedge_{t=0}^k \pi(t)(r) = a_0,  \vspace*{2mm} \ \\
     \$(r) & \ \hbox{ otherwise. } 
     \end{array} \right.
\]
and the cost of a path $\pi \in \Pi(M)$ with regard to all resources in $Res$ is defined as
\[
c_{Res}(\pi) \ = \ \sum_{r \in Res} c_r(\pi)
\]
where $\pi(t)$ denotes the $t$-th state of $\pi$.
\end{definition}

  \noindent
  Hence,  the resource costs of a path $\pi$ is the sum of prices of the resources that are ever allocated by some agent in the states along $\pi$.  Conversely, resources that are never allocated do not contribute to the costs. 
We consider MRAs that run until all goals have been achieved where each goal has a deadline.  Thus, in the calculation of the costs of a path it is sufficient to only take the $k$-prefix of the path into account where $k$ is the latest deadline.  Based on path costs we can now define what a resource cost-optimal winning strategy is.

  \begin{definition}[Resource Cost-Optimal Strategy]
  Let $M = (Agt, T,Res,\$, \Phi)$ be an MRA, let $s_0$ be its initial state, and let $\sigma_{Agt}$ be a joint strategy. 
  Then $\sigma_{Agt}$ is a cost-optimal winning strategy with regard to resources if the following conditions hold: 
  \begin{enumerate}
  \item $[M,\pi(s_0, \sigma_{Agt}) \models  \Phi]$ \ \\
(winning)  
   \vspace*{2mm}
  \item $\forall \sigma'_{Agt} \neq \sigma_{Agt} : [M,\pi(s_0, \sigma'_{Agt}) \models  \Phi] \ \rightarrow \ c_{Res}(\pi(s_0, \sigma'_{Agt})) \geq c_{Res}(\pi(s_0, \sigma_{Agt}))$ \ \\
  (resource cost-optimal) 
  \end{enumerate}
 \end{definition}

 \noindent
Hence,  a winning strategy $\sigma_{Agt}$ is cost-optimal
with regard to resources if following this strategy results in a path whose costs are less or equal to the costs of the paths resulting from any other winning strategy $\sigma'_{Agt}$. 
We denote the corresponding cost-optimal strategy synthesis problem by $[M,s \models \coop{Agt}\Phi]_{Opt}^{Res}$.

The actions listed in Table 2 characterise a resource cost-optimal winning strategy for the example system $M_{ex}$. The strategy makes use of the resources $r_1, r_2,r_3 $ and $r_5$ whereas the resources $r_4$ and $r_6$ are never used. According to the price function of $M_{ex}$,  the cost of the path resulting from this strategy is 7.  In contrast, the cost of the path resulting from the non-optimal winning strategy from Table 1 is 10.  This strategy additionally uses the resource $r_6$. 

\begin{table}[h]
\caption{Actions that characterise a resource cost-optimal strategy.}
\begin{center}
\begin{tabular}{r|c|c|c|c|c}
\hline
 time step: & 0 & 1 & 2 & 3 & 4 \\ \hline
  $a_1$'s actions: & $req_{r_5}$ & $rel_{all}$ & $req_{r_1}$ & $req_{r_3}$ & $rel_{all}$ \\ \hline
  $a_2$'s actions: & $idle$ & $req_{r_2}$ & $req_{r_5}$ & $rel_{all}$ & $idle$  \\ \hline
   $a_3$'s actions: & $req_{r_3}$ & $rel_{all}$ & $idle$ & $idle$ & $idle$ \\ \hline
\end{tabular}
\end{center}
\end{table}
\noindent
Assuming that our example MRA models a resource allocation problem in distributed computing,  the resource cost-optimal strategy gives us an indication which types and amounts of resources actually need to be purchased for an agent-based solution that guarantees the achievement of all resource goals.  
So far, we assumed that the number of agents in the system is fixed and that goals are pre-assigned to agents. 
In the following we will relax this condition and introduce a generalised cost-optimal strategy synthesis that also considers costs for agents.

   \subsection{MRA Cost-Optimal Strategy Synthesis}
For our generalised  cost-optimal strategy synthesis we define agent costs of paths.

  \begin{definition}[Agent Costs of Paths]
Let $M = (Agt, T,Res,\$, \Phi)$ be an MRA, 
and let $k = \textbf{\emph{max}}(d | (R,p,d)\in \Phi)$ be the latest deadline. 
Moreover, let $\$_{A} \in \mathbb{N}$ be the price of each agent. 
The cost of a path $\pi \in \Pi(M)$ with regard to an agent $a \in Agt$ is defined as 
\[
c_a(\pi) \ = \ \left\{ \begin{array}{ll} 
     0 & \ \hbox{ if } \ \bigwedge_{t=0}^k \bigwedge_{r \in Res} \pi(t)(r) \neq a,  \vspace*{2mm} \ \\
     \$_A & \ \hbox{ otherwise. } 
     \end{array} \right.
\]
and the cost of a path $\pi \in \Pi(M)$ with regard to all agents in $Agt$ is defined as
\[
c_{Agt}(\pi) \ = \ \sum_{a \in Agt} c_a(\pi)
\]
where $\pi(t)$ denotes the $t$-th state of $\pi$.
\end{definition}

 \noindent
 Hence, the agent costs of a path is the sum of the prices of the agents that ever allocate some resources along the path. 
 We can now straightforwardly define the overall costs of paths with regard to resources and agents. 

 \begin{definition}[MRA Costs of Paths]
Let $M = (Agt, T,Res,\$, \Phi)$ be an MRA, 
and let $k = \textbf{\emph{max}}(d | (R,p,d)\in \Phi)$ be the latest deadline. 
Moreover, let $\$_{A} \in \mathbb{N}$ be the price of each agent. 
The cost of a path $\pi \in \Pi(M)$ with regard to $M$ is defined as 
\[
c_{M}(\pi) \ = \ c_{Res}(\pi) + c_{Agt}(\pi).
\]
\end{definition}

\noindent
This allows us to consider a generalised scenario where the goals to be achieved are not assigned to particular agents. 

\begin{definition}[MRA Model Checking of General Goal Properties]
Let $M = (Agt, T,Res,\$, \Phi^*)$ be an MRA
where the special set $\Phi^* = \{\phi^1, \ldots, \phi^g\}$ consists of goals that are not associated with particular agents.   
 Then the strategic MRA model checking problem $[M,s \models \coop{Agt}\Phi^*]$ is inductively  defined
  as follows:        
  \[
  \begin{array}{lcllcl}
    
    [M,s \models \coop{Agt}\Phi^*] \ & \ \equiv \ & \ \bigvee_{\sigma_{Agt} \in \Sigma_{Agt}} \ [M,\pi(s, \sigma_{Agt}) \models  \Phi^*]    \vspace*{0.75mm} \ \\ 
    
    
        [M,\pi \models \Phi^*] \ & \ \equiv \ & \
\bigwedge_{\phi \in \Phi^*}    \       [M,\pi  \models  \phi]    \vspace*{0.75mm} \ \\ 


         [M,\pi \models \phi] \ & \ \equiv \ & \
\bigvee_{a \in Agt} [M,\pi \models \phi_a]
    \vspace*{0.75mm} \ \\ 
    
  \end{array}
  \]
  where $[M,\pi \models \phi_a]$ is defined according to Definition 9.  
  \end{definition}
  
  \noindent
  Thus,  model checking of general goal properties searches for a strategy that ensures that each goal will be achieved by \emph{some} agent.  In this context, a cost-optimal winning strategy is one that minimises the combined costs of required resources \emph{and} agents. 
  
    \begin{definition}[MRA Cost-Optimal Strategy]
  Let $M = (Agt, T,Res,\$, \Phi^*)$ be an MRA where the special set $\Phi^* = \{\phi^1, \ldots, \phi^g\}$ consists of goals that are not associated with particular agents.   
  Moreover,  let $s_0$ be the initial state of $M$, let $\$_A$ be the price of each agent, and let $\sigma_{Agt}$ be a joint strategy. 
  Then $\sigma_{Agt}$ is a cost-optimal winning strategy with regard to $M$ if the following conditions hold: 
  \begin{enumerate}
  \item $[M,\pi(s_0, \sigma_{Agt}) \models  \Phi^*]$ \ \\
  (winning) \vspace*{2mm}
  \item $\forall \sigma'_{Agt} \neq \sigma_{Agt} : [M,\pi(s_0, \sigma'_{Agt}) \models  \Phi^*] \ \rightarrow \ c_{M}(\pi(s_0, \sigma'_{Agt})) \geq c_{M}(\pi(s_0, \sigma_{Agt}))$ \ \\
  (MRA cost-optimal)
  \end{enumerate}
 \end{definition}
 
 
  \noindent
Hence,  a winning strategy $\sigma_{Agt}$ is MRA cost-optimal
if following this strategy results in a path whose combined resource and agent costs are less or equal to the costs of the paths resulting from any other winning strategy $\sigma'_{Agt}$. 
We denote the corresponding cost-optimal strategy synthesis problem by $[M,s \models \coop{Agt}\Phi]_{Opt}^{M}$.

Let us now consider a slight variant of our example system $M_{ex}$ where the goals are no longer pre-assigned to agents.  The actions listed in Table 3 characterise an MRA cost-optimal winning strategy for this variant.  The resource costs are still the same as in the strategy depicted in Table 2.  But the MRA cost-optimal strategy requires just  two instead of three agents for achieving all goals.  The role of agent $a_3$ is insignificant here.  Thus, the strategy indicates that only two agents need to be `purchased'.

 \begin{table}[h]
\caption{Actions that characterise a MRA cost-optimal strategy.}
\begin{center}
\begin{tabular}{r|c|c|c|c|c}
\hline
 time step: & 0 & 1 & 2 & 3 & 4 \\ \hline
  $a_1$'s actions: & $req_{r_5}$ & $rel_{all}$ & $req_{r_1}$ & $req_{r_3}$ & $rel_{all}$ \\ \hline
  $a_2$'s actions: & $req_{r_3}$ & $rel_{all}$ & $req_{r_2}$ & $req_{r_5}$ & $rel_{all}$    \\ \hline
   $a_3$'s actions: & $idle$  &  $idle$ &  $idle$& $idle$ & $idle$ \\ \hline
\end{tabular}
\end{center}
\end{table}
 
  \section{Reduction of Cost-Optimal Strategy Synthesis to Weighted Max-SAT}
In this section we show how cost-optimal strategy synthesis  can be reduced to weighted maximum satisfiability solving. 
  
 \subsection{Weighted Maximum Satisfiability Problem}
 The weighted maximum satisfiability problem is a generalisation of the Boolean satisfiability problem of propositional logic formulas in conjunctive normal form. 
 
 \begin{definition}[Conjunctive Normal Form (CNF)]
Let $Var$ be a set of Boolean variables.
A propositional logic formula $\mathcal{F}$ over $Var$ in \emph{conjunctive normal form} is a conjunction of clauses $\mathcal{C}$ where each clause is a disjunction of literals $l$, and a literal is either a variable $v \in Var$ or its negation $\neg v$.
\end{definition}

\noindent
For CNF formulas the satisfiability problem is defined as follows:
\begin{definition}[Boolean Satisfiability Problem]
Let $\mathcal{F}$ over $Var$ be a formula in conjunctive normal form.  The Boolean satisfiability problem with regard to $\mathcal{F}$ is the problem of determining whether there exists a truth assignment $\alpha : Var \rightarrow \{\textbf{{\emph{0}}},\textbf{{\emph{1}}}\}$ that makes all clauses of $\mathcal{F}$ true.  Boolean satisfiability can be also defined as a function
\[
\textbf{\emph{sat}}(\mathcal{F}) \ = \ \left\{ \begin{array}{ll} 
     \textbf{\emph{1}} & \ \hbox{ if } \exists \alpha \in \mathcal{A}(Var) \hbox{ with } \alpha(\mathcal{F}) = \textbf{\emph{1}},
\vspace*{2mm} \ \\
     \textbf{\emph{0}} & \ \hbox{ otherwise,} 
     \end{array} \right.
\]
where $\mathcal{A}(Var)$ is the set of all possible truth assignments over $Var$. 
\end{definition}
%
%

\noindent
Weighted conjunctive normal form extends CNF by assigning non-negative weights to each clause of a formula. 

\begin{definition}[Weighted Conjunctive Normal Form (WCNF)]
Let $Var$ be a set of Boolean variables.
A propositional logic formula $\mathcal{F}$ over $Var$ in weighted conjunctive normal form is a conjunction of weighted clauses $(\mathcal{C}, w_{\mathcal{C}})$ where $\mathcal{C}$ is a standard clause and $w_{\mathcal{C}} \in \mathbb{N}_{\infty}$ is its weight. 
A clause $(\mathcal{C}, w_{\mathcal{C}})$ with $w_{\mathcal{C}} \in \mathbb{N}$ is called a soft clause and a clause $(\mathcal{C}, \infty)$ is called a hard clause.
\end{definition}

\noindent
For the sake of simplicity we typically just write $\mathcal{C}$ for hard clauses $(\mathcal{C}, \infty)$.  Each WCNF formula $\mathcal{F}$ can be written as a conjunction $\mathcal{H} \wedge \mathcal{S}$ where $\mathcal{H}$ are the hard clauses and $\mathcal{S}$ are the soft clauses of $\mathcal{F}$.
For WCNF formulas the following optimisation problem has been defined:

\begin{definition}[Weighted Maximum Satisfiability Problem] 
Let $\mathcal{F} = \mathcal{H} \wedge \mathcal{S}$ over $Var$ be a propositional logic formula in weighted conjunctive normal form where $\mathcal{H}$ are the hard clauses and $\mathcal{S}$ are the soft clauses.  The weighted maximum satisfiability problem with regard to $\mathcal{F}$ is the problem of finding a truth assignment $\alpha : Var \rightarrow \{\textbf{{\emph{0}}},\textbf{{\emph{1}}}\}$ that
maximises
\[
\sum\limits_{(\mathcal{C}, w_{\mathcal{C}}) \in \mathcal{S}} \alpha(\mathcal{C}) \cdot w_{\mathcal{C}} 
\]
subject to the condition that $\alpha(\mathcal{H}) = \textbf{{\emph{1}}}$  holds. 
Weighted maximum satisfiability can be defined as a function
\[
\hspace*{-7.5mm}
\textbf{\emph{max-sat}}(\mathcal{F}) \ = \ \left\{ \begin{array}{ll} 
     nil & \ \hbox{ if } \textbf{\emph{sat}}(\mathcal{H}) = \textbf{\emph{0}},
\vspace*{2mm} \ \\
     \underset{\alpha \in \mathcal{A}({Var})}{\operatorname{\textbf{\emph{arg max}}}}  \bigg(\alpha(\mathcal{H}) \cdot \big( \sum\limits_{(\mathcal{C}, w_{\mathcal{C}}) \in \mathcal{S}} \alpha(\mathcal{C}) \cdot w_{\mathcal{C}} \big) \bigg) & \ \hbox{ otherwise, } 
     \end{array} \right.
\]
where $\mathcal{A}(Var)$ is the set of all possible truth assignments over $Var$. 
\end{definition}

\noindent
Hence, the solution of the weighted maximum satisfiability problem with regard to $\mathcal{F}$ is a truth assignment $\alpha$ that \emph{maximises} the sum of weights of the satisfied soft clauses, under the condition that \emph{all} hard clauses are satisfied. If no such assignment exists, then the weighted maximum satisfiability problem has no solution. 
We will now show how cost-optimal strategy synthesis problems can be encoded as weighted Max-SAT problems.

 \subsection{Max-SAT Encoding of Resource Cost-Optimal Strategy Synthesis}
In \cite{timm2021model} we showed how to encode standard MRA strategy synthesis problems $[M,s_0 \models \coop{Agt} \Phi]$ as propositional logic formulas $\mathcal{F}$ such that the following equivalence holds:
 \[
 [M,s_0 \models \coop{Agt} \Phi] \ \equiv \ \textbf{sat}\big(\mathcal{F}\big)
 \]
Hence, the synthesis of winning strategies can be performed via satisfiability solving. 
A further property of our encoding is that each truth assignment $\alpha$ that satisfies $\mathcal{F}$ characterises a winning strategy $\sigma_{Agt}^{\alpha}$. 
The overall encoding is a conjunction $\mathcal{F} =  [\coop{Agt}] \wedge [M] \wedge   [\Phi]$ 
where $[\coop{Agt}]$ encodes that all agents must follow a uniform strategy and adhere to the protocol,  $[M]$ encodes the feasible paths of $M$,  and $[\Phi]$ restricts the paths to those that satisfy all goals in $\Phi$.  
In  \cite{timm2021model} we only considered simple goals, one for each agent and  with regard to a single type of resource.
Thus, for the more complex goals that we focus on in this work, we introduce the following extended encoding of goals:

%
%
 
 \begin{definition}[Encoding of Goals] Let $M = (Agt, T,Res,\$, \Phi)$ be an MRA.
Then the agent goal property $\Phi$ is encoded in propositional logic as
 \[
  [\Phi] =  \bigwedge_{\phi_a \in \Phi} \bigvee_{t=0}^{d-p} \bigwedge_{\tau \in R} \bigvee_{r \in Res_{\tau}} \bigwedge_{t'=t}^{t+p} \ [r=a]_t
 \]
 Let $M = (Agt, T,Res,\$, \Phi^*)$ be an MRA.
Then the general goal property $\Phi^*$ is encoded in propositional logic as
   \[
  [\Phi^*] =  \bigwedge_{\phi \in \Phi} \bigvee_{a \in Agt}\bigvee_{t=0}^{d-p} \bigwedge_{\tau \in R} \bigvee_{r \in Res_{\tau}} \bigwedge_{t'=t}^{t+p} \ [r=a]_t
 \]
  where $\phi_a$ resp.  $\phi$ is assumed to be the tuple $(R,p,d)$, 
and $[r=a]_t$ encodes that resource $r$ is allocated by agent $a$ at time step $t$.
\end{definition}

\noindent
The definition of the sub encoding $[r=a]_t$ can be found in \cite{timm2021model}. 

In order to enable the synthesis of \emph{resource cost-optimal} winning strategies, we further extend our encoding.  
Solving this optimisation problem involves to find a strategy that minimises the costs of the resources that are actually used.  We can equivalently search for a strategy that maximises the costs of the unused resources, which makes the problem compatible with Max-SAT. 
The first part of the extension introduces auxiliary variables $nu_r$ that encode that a resource $r \in Res$ is never used.
 
\begin{definition}[Auxiliary Encoding -- Resource Costs] 
Let $M = (Agt, T,Res,\$, \Phi)$ be an MRA and let $k = \textbf{\emph{max}}(d | (R,p,d)\in \Phi)$ be the latest deadline.
Then the auxiliary encoding for resource cost optimisation is
\[
[Aux_{Res}] = \bigwedge_{r \in Res}  \big( nu_r \leftrightarrow  (\bigwedge_{t=0}^k [r=a_0]_t)  \big)
\]
where $nu_r$ with $r \in R$ are the auxiliary variables introduced for the encoding, and $[r=a_0]_t$ encodes that resource $r$ is unallocated at time step $t$.
\end{definition} 

\noindent
If we conjunctively add the auxiliary encoding $[Aux_{Res}]$ to the overall encoding of a strategy synthesis problem,  this will not affect the satisfiability.
But now we have that a truth assignment $\alpha$ will set an auxiliary variable $nu_r$ to \emph{true} if and only if $\alpha$ characterises a  strategy $\sigma_{Agt}^{\alpha}$ and the resource $r$ will be never used when the strategy $\sigma_{Agt}^{\alpha}$ is followed. 
Hence,  $nu_r$ is a single-variable encoding of $r$ never being used.  
We can now utilise each $nu_r$ as a unit clause in the optimisation extension of our encoding:

\begin{definition}[Resource Costs Optimisation Encoding] 
Let $M = (Agt, T,Res,\$, \Phi)$ be an MRA. 
Then the resource costs optimisation encoding is
\[
[Opt_{Res}] = \bigwedge_{r \in Res} (nu_r, \$(r))
\]
where $nu_r$ with $r \in Res$  are the Boolean variables introduced in the  auxiliary encoding for resource cost optimisation.  
\end{definition}

\noindent
 Hence,  $[Opt_{Res}]$ consists of soft clauses $(nu_r, \$(r))$, one for each $r \in Res$, 
where $nu_r$ encodes  that $r$ is never used 
 and $\$(r)$ (the price of the resource $r$) is the weight of the clause. 
The overall encoding of the problem of synthesising a resource cost-optimal strategy  is $\mathcal{F}_{Opt}^{Res} = [\coop{Agt}] \wedge [M] \wedge   [\Phi] \wedge [Aux_{Res}] \wedge [Opt_{Res}]$
where the sub formula $[\coop{Agt}] \wedge [M] \wedge   [\Phi] \wedge [Aux_{Res}]$ consists of  hard clauses only.  Solving \textbf{max-sat}$(\mathcal{F}_{Opt}^{Res})$ will return a truth assignment that satisfies all hard clauses and maximises the sum of weights of satisfied soft clauses, if such an assignment exists.  Such an assignment characterises a winning strategy that is resource cost-optimal. 

 \begin{theorem}[Resource Cost-Optimal Strategy Synthesis] 
Let $[M,s_0 \models  \Phi]_{Opt}^{Res}$ be a resource cost-optimal strategy synthesis problem and let $\mathcal{F}^{Res}_{Opt}$ be its WCNF encoding.
Then the following properties hold:
\begin{enumerate}
\item
If \textbf{{\emph{max-sat}}}$(\mathcal{F}^{Res}_{Opt}) = nil$, then there does not exist a joint winning strategy for achieving all goals in $\Phi$ with the resources in $Res$.
\vspace*{2mm}
\item
If \textbf{{\emph{max-sat}}}$(\mathcal{F}^{Res}_{Opt}) = \alpha$,  then $\alpha$ characterises a resource cost-optimal winning strategy $\sigma_{Agt}^{\alpha}$.
%
\end{enumerate}

\end{theorem}
\medskip
\noindent
\emph{Proof.} \ \\
We have that $
\mathcal{F}^{Res}_{Opt}   =   \mathcal{H}^{Res}_{Opt} \wedge \mathcal{S}^{Res}_{Opt}
$
where
$
\mathcal{H}^{Res}_{Opt}   =   [\coop{Agt}] \wedge [M] \wedge   [\Phi] \wedge [Aux_{Res}]
$
are the hard clauses and
$
\mathcal{S}^{Res}_{Opt}   =     [Opt_{Res}]
$
are the soft clauses of the encoding. 
Moreover, the WCNF formula $\mathcal{F}^{Res}_{Opt}$ is defined over a set of Boolean variables $Var$, and $\mathcal{A}(Var)$ is the set of all possible truth assignments over $Var$.

\medskip
\noindent
\emph{Proof of Property 1:} \ \\
\textbf{{max-sat}}$(\mathcal{F}^{Res}_{Opt}) = nil$ implies  that \textbf{{sat}}$([\coop{Agt}]   \wedge [M] \wedge [\Phi]  \wedge [Aux_{Res}]) =$ \textbf{0} (Definition 19).  The sub formula
$[Aux_{Res}]$  only introduces auxiliary variables and sets them equivalent to resource properties (Definition 21).  Hence,  $[Aux_{Res}]$ cannot be the cause of unsatisfiability and we can conclude that \textbf{{sat}}$([\coop{Agt}]   \wedge [M] \wedge [\Phi]) =$ \textbf{0}.  This implies that $[M,s_0 \models \coop{Agt} \Phi ]$ does not hold (shown in \cite{timm2021model}).  We can now immediately conclude that there does not exist a joint winning strategy to achieve $\Phi$.

\medskip
\noindent
\emph{Proof of Property 2:} \ \\
\textbf{{max-sat}}$(\mathcal{F}^{Res}_{Opt}) = \alpha$ implies that
$
\alpha  =  \underset{\alpha \in \mathcal{A}({Var})}{\operatorname{\textbf{arg max}}}  \bigg(\alpha(\mathcal{H}^{Res}_{Opt}) \cdot \big( \sum\limits_{(\mathcal{C}, w_{\mathcal{C}}) \in \mathcal{S}^{Res}_{Opt}} \alpha(\mathcal{C}) \cdot w_{\mathcal{C}} \big) \bigg)
$
(Definition 19).  Hence, the  assignment $\alpha$ satisfies all hard clauses of $\mathcal{F}^{Res}_{Opt}$.  In particular, $\alpha([\coop{Agt}]   \wedge [M] \wedge [\Phi]) =$ \textbf{1}. We can conclude that $[M,s_0 \models \coop{Agt} \Phi ]$ holds, and that $\alpha$ characterises a corresponding winning strategy $\sigma_{Agt}^{\alpha}$ (shown in \cite{timm2021model}). 
Now we still need to prove that $\sigma_{Agt}^{\alpha}$ is cost-optimal with regard to the use of resources.  
According to the definition of \textbf{{max-sat}},  out of all truth assignments that satisfy all hard clauses, $\alpha$ is the  assignment that maximises the sum of weights of the satisfied soft clauses.  Let $w^s_{\alpha}$ be the sum of weights of the  soft clauses satisfied by $\alpha$.  Then for all other truth assignments $\alpha'$ that satisfy all hard clauses we have that $w^s_{\alpha} \geq w^s_{\alpha'}$ (Definition 19).  
Conversely, let $w^u_{\alpha}$ be the sum of weights of the  soft clauses \emph{not} satisfied by $\alpha$.  Then for all other truth assignments $\alpha'$ that satisfy all hard clauses we have that $w^u_{\alpha} \leq w^u_{\alpha'}$.
Each soft clause is of the form $(nu_r, \$(r))$.
For a resource $r \in Res$,  $\alpha(nu_r) =$ \textbf{1} implies that $\alpha(\bigwedge_{t=0}^k [r=a_0]_t) =$ \textbf{1} (Definition 21).  
The hard clauses of the encoding, $[\coop{Agt}] \wedge [M] \wedge   [\Phi]$, ensure that $\alpha(\bigwedge_{t=0}^k [r=a_0]_t) =$ \textbf{1} if and only if $\alpha$ characterises a joint strategy $\sigma^{\alpha}_{Agt}$ that results in a path $\pi(s_0,\sigma^{\alpha}_{Agt})$ where resource $r$ is never allocated by any agent at all time steps until $k$ where $k$ is the latest deadline (shown in \cite{timm2021model}).
We can conclude that $\alpha$ characterises a strategy that results in a $k$-bounded path where $w^s_{\alpha}$ are the cumulative costs of the resources that are never used,  and $w^u_{\alpha}$ are the cumulative costs of the resources that are  actually used.  Thus, $c_{Res}(\pi(s_0,\sigma^{\alpha}_{Agt})) = w^u_{\alpha}$ (Definition 10).  
Based on a similar argumentation, we can show that each alternative truth assignment $\alpha'$ that satisfies all hard clauses characterises a path $\pi(s_0,\sigma^{\alpha'}_{Agt})$
with cumulative costs of resources $c_{Res}(\pi(s_0,\sigma^{\alpha'}_{Agt})) = w^u_{\alpha'}$.
We already argued that $w^u_{\alpha} \leq w^u_{\alpha'}$ for all alternative assignments $\alpha'$. Hence, we can conclude that $\alpha$ characterises a joint winning strategy $\sigma^{\alpha}_{Agt}$ that is resource cost-optimal (Definition 11).

\noindent
\qed

\medskip
\noindent
This theorem can be utilised as follows: 
Several practically relevant resource allocation problems in distributed computing can be modelled as an MRA $M$ where the agents represent the computers of the distributed system.  While it is of primary importance that the system will achieve all goals, it is of secondary importance that the costs of the resources that need to be purchased are as low as possible.  In the MRA model of the distributed system,  we can initialise $Res$ as a saturated set 
$\{r_{\tau} | \forall (R,p,d) \in \Phi \ \forall \tau \in R \}$
that contains a resource for each type of each resource composition of each goal. 
The resource cost-optimal strategy synthesis problem with regard to $M$ can then be encoded in propositional logic.
Max-SAT-based solving of the encoded problem 
will not only yield an optimal strategy, it will also indicate which resources of the model are actually dispensable.  We can safely remove these resources from $Res$, and it is still guaranteed that all goals can be achieved.  
For the modelled distributed system only the indispensable resources need to be purchased.  Hence, our Max-SAT-based technique can be used to synthesize cost-optimal MRA solutions to resource allocation problems.  

In the next sub section we show that our Max-SAT approach can be straightforwardly extended to the optimisation of systems where also agents need to be purchased and where goals can be assigned to arbitrary agents. 

 \subsection{Max-SAT Encoding of MRA Cost-Optimal Strategy Synthesis}
 For the Max-SAT encoding of MRA cost-optimal strategy synthesis problems we need to introduce further auxiliary variables, one for each agent in the system.

\begin{definition}[Auxiliary Encoding -- Agent Costs] 
Let $M = (Agt, T,Res,\$, \Phi^*)$ be an MRA and let $k = \textbf{\emph{max}}(d | (R,p,d)\in \Phi^*)$ be the latest deadline.
Then the auxiliary encoding for agent cost optimisation is
\[
[Aux_{Agt}] = \bigwedge_{a \in Agt} \big(  nu_a \leftrightarrow (\bigwedge_{t=0}^k \bigwedge_{Res} \neg [r=a]_t)  \big)
\]
where $nu_a$ with $a \in Agt$ are the auxiliary variables introduced for the encoding,  and $[r=a]_t$ encodes that resource $r$ is allocated by agent $a$ at time step $t$.
\end{definition}

\noindent
Hence, each $nu_a$ encodes that agent $a$ never allocates any resource.  We now utilise each  $nu_a$ as a unit clause in the optimisation of agent costs encoding.

\begin{definition}[Agent Costs Optimisation Encoding] 
Let $M = (Agt, T,Res,\$, \Phi^*)$ be an MRA and let $\$_{A} \in \mathbb{N}$ be the price of each agent. 
Then the agent costs optimisation encoding is
\[
[Opt_{Agt}] = \bigwedge_{a \in Agt} (nu_a, \$_A)
\]
where $nu_a$ with $a \in Agt$  are the Boolean variables introduced in the  auxiliary encoding for agent cost optimisation.  
\end{definition}

\noindent
 Hence,  $[Opt_{Agt}]$ consists of soft clauses $(nu_a, \$_A)$, one for each $a \in Agt$, 
where $nu_a$ encodes  that $a$ is never used 
 and $\$_A$ is the weight of the clause. 
The overall encoding of the problem of synthesising an MRA cost-optimal strategy  is $\mathcal{F}_{Opt}^{M} = \mathcal{F}_{Opt}^{Res}   \wedge [Aux_{Agt}] \wedge [Opt_{Agt}]$.
Solving \textbf{max-sat}$(\mathcal{F}_{Opt}^{M})$ will return a truth assignment that satisfies all hard clauses and maximises the sum of weights of satisfied soft clauses, if such an assignment exists.  Such an assignment characterises a winning strategy that is MRA cost-optimal. 

 \begin{theorem}[MRA Cost-Optimal Strategy Synthesis] 
Let $[M,s_0 \models  \Phi^*]_{Opt}^{M}$ be an MRA cost-optimal strategy synthesis problem and let $\mathcal{F}^{M}_{Opt}$ be its WCNF encoding.
Then the following properties hold:
\begin{enumerate}
\item
If \textbf{{\emph{max-sat}}}$(\mathcal{F}^{M}_{Opt}) = nil$,  
then there does not exist a joint winning strategy for achieving all goals in $\Phi^*$ with the resources $Res$ and agents $Agt$.
%
\vspace*{2mm}
\item
If \textbf{{\emph{max-sat}}}$(\mathcal{F}^{M}_{Opt}) = \alpha$,  then $\alpha$ characterises an MRA cost-optimal winning strategy $\sigma_{Agt}^{\alpha}$.
%
\end{enumerate}

\end{theorem}

\noindent
The proof of Theorem 2 is analogous to the proof of Theorem 1.  

\medskip
\noindent
Thus, given a set of resource allocation goals $\Phi^*$ this theorem can be exploited to synthesise a winning strategy that does not only minimise the costs of required resources but also results in an optimal assignment of goals to agents. 
Such an optimal assignment will ensure that the resource allocation problem associated with $\Phi^*$ will be solved with a cost-minimal amount of agents and resources.

\section{Implementation and Experiments}
We developed the tool SATMAS\footnote{available at \url{https://github.com/TuksModelChecking/Satmas/tree/nur_optimization}} that implements our synthesis approach in Python.  SATMAS takes a specification of a multi-agent system for resource allocation $M = (Agt, T,Res,\$, \Phi)$ as an input.  The output is a resource cost-optimal winning strategy $\sigma^{\alpha}_{Agt}$ and a corresponding cost-optimal set of required resources if existent. Otherwise, the tool outputs `\emph{no winning strategy exists}'.
SATMAS encodes the synthesis problem in propositional logic. 
The maximum-satisfiability solver OPEN-WBO \cite{martins2021open} is employed to determine a truth assignment for the encoding that satisfies all hard clauses and maximises the sum of weights of the satisfied soft clauses.  From such an assignment the corresponding cost-optimal strategy and the cost-optimal set of required resources can be immediately derived.  The extension of our tool to the synthesis of strategies that are additionally optimal with regard to agent costs is in preparation. 

In experiments we were able to synthesise resource cost-optimal strategies for MRAs consisting of ten agents and six different types of resources in less than a minute.
Table 4 shows our experimental results.
The \emph{Scenario} column indicates the number of agents $Agt = \{a_1,\ldots, a_n\}$ and the number of resource types $T = \{\tau_1,\ldots, \tau_m\}$ in the MRA.  As a price function we used $\$(\tau_i)=i$.
For simplicity, we fixed the number of goals of each agent as well as the period of each goal to one. 
The resource composition of each goal was randomly selected under the constraint that the total number of resource types associated with a goal is from the interval $\mathcal{R}$.  Moreover, the deadline of each goal was randomly selected from the interval $\mathcal{D}$. 
The column \emph{Costs} indicates the total costs of resources associated with the synthesised cost-optimal winning strategy.  The column \emph{Time} shows the overall time that the tool spent on encoding and satisfiability solving.  The experiments were conducted on a 2.6 GHz Intel Core i7 system with 16 GB.

 \begin{table}[h]
\begin{center}
 \caption{Cost-Optimal Strategy Synthesis.} 
	\setlength{\tabcolsep}{0.6em}
	{\renewcommand{\arraystretch}{1.15}
	\begin{tabular}{c|c|c}
	\hline
\ \ 	Scenario \ \ &    \ \ Costs \ \ &  \ \ Time \ \ \\
		\hline
	$\vert Agt \vert = 2$,     $\vert T \vert = 3$, $\mathcal{R} = [1,2]$,  $\mathcal{D} = [5,15]$ &   3   & $0.2s$  \\
			\hline
				$\vert Agt \vert = 4$,     $\vert T \vert = 4$, $\mathcal{R} = [1,3]$,  $\mathcal{D} = [5,15]$ &   6  & $1.0s$  \\
			\hline
				$\vert Agt \vert = 6$,    $\vert T \vert = 4$, $\mathcal{R} = [2,3]$,  $\mathcal{D} = [5,20]$ &   6 & $3.5s$  \\
			\hline
							$\vert Agt \vert = 8$,    $\vert T \vert = 6$, $\mathcal{R} = [2,5]$,  $\mathcal{D} = [5,30]$  & 15 & $20.4s$  \\
			\hline
				$\vert Agt \vert = 10$,    $\vert T \vert = 6$, $\mathcal{R} = [2,4]$,  $\mathcal{D} = [5,30]$ &   10   & $54.3s$  \\
			\hline
	\end{tabular}
	}
\end{center}
\end{table}

\noindent
Thus, for our randomly generated MRA consisting of ten agents and six types of resources there exists a cost-optimal winning strategy that involves resource costs of 10.  We ran the synthesis again for the same MRA but with having the optimisation disabled. This resulted in a winning strategy with resource costs of 21.  Hence,  our optimisation approach can enable significant savings in terms of resource costs. 

\section{Related Work}
Model checking and strategy synthesis for multi-agent systems has been originally proposed in \cite{alur2002alternating}.  The authors introduce the alternating-time logics ATL and ATL$^*$,  which are logics for reasoning about strategic abilities of agents. 
The general ATL model checking problem is PTIME-complete whereas the ATL$^*$ model checking problem is 2EXPTIME-complete.  Thus, while for ATL model checking BDD-based tools like MCMAS \cite{lomuscio2017mcmas} and MOCHA \cite{alur1998mocha} exist, ATL$^*$ has been rather considered on a theoretical level \cite{schewe2008atl}. 
An existing tool for synthesising ATL strategies is SMC \cite{pilecki2017smc}, which operates on a BDD model of the multi-agent system to be verified.  It iteratively guesses a strategy, fixes the strategy in the model and checks whether it is a winning strategy, which reduces ATL model checking to CTL model checking \cite{cimatti2000nusmv} in each iteration.  
The techniques and tools mentioned above focus on synthesising winning strategies for \emph{general} multi-agent systems and alternating-time properties. 
In contrast, our approach focusses on systems and properties with regard to \emph{resource allocation problems} \cite{nair2018multi}.  Moreover, we aim at synthesising winning strategies that are additionally \emph{optimal} in terms of costs for resources and agents. 

Our approach is related to strategy synthesis for systems with combined qualitative and quantitative objectives \cite{chatterjee2005mean,chatterjee2012energy,gutierrez2017nash,gutierrez2021equilibria,bulling2022combining}. 
In \cite{chatterjee2005mean,chatterjee2012energy} parity games are studied where the objectives combine qualitative $\omega$-regular requirements and quantitative energy usage requirements. 
The authors of \cite{gutierrez2017nash,gutierrez2021equilibria} introduce multi-agent systems in which agents have a primary objective that is qualitative and a secondary objective that is quantitative.  In \cite{gutierrez2021equilibria} it is shown that if the qualitative objectives are LTL formulas then the problem is 2EXPTIME-complete.  
Another framework for reasoning about systems where agents have both qualitative and quantitative objectives is proposed in \cite{bulling2022combining}.  The authors augment multi-agent systems with transition pay-offs and introduce a quantitative extension of the logic ATL$^*$.  This allows to model check whether agents have the strategic abilities to achieve quantitative pay-off objectives and at the same time qualitative state-based objectives.   
Our approach is similar to the above in the sense that we consider systems where resource allocation goals are qualitative objectives and cost-optimality is a quantitative objective.  However,  in \cite{gutierrez2017nash,gutierrez2021equilibria, bulling2022combining} it is assumed that each agent has individual quantitative objectives.  Hence,  these works aim at synthesising Nash equilibrium strategies \cite{nash1950equilibrium} rather than overall optimal strategies.  
Moreover,  the above works are predominantly of theoretical nature and focus on establishing general complexity results.  In contrast, our work considers a more specific resource allocation scenario and we propose a practical approach to solve the synthesis problem, which is based on a Boolean encoding and on maximum satisfiability solving (Max-SAT) \cite{hansen1990algorithms}. 

To the best of our knowledge,  our technique is the first Max-SAT-based approach to the synthesis of cost-optimal strategies for multi-agent systems. 
In related fields,  Max-SAT has been employed to find optimal coalitions of agents for solving network security problems \cite{2014maximum},  to synthesise reactive  controllers of autonomous systems \cite{dimitrova2020reactive},  and to model check quantitative hyper-properties \cite{finkbeiner2018model}.

\section{Conclusion and Outlook}
In this paper, we presented a Max-SAT-based technique for synthesising winning strategies for multi-agent systems for resource allocation.  The synthesised strategies do not only ensure that all resource allocation goals will be achieved, they also result in a cost-optimal use of resources and agents.  Our approach can be utilised to model resource allocation problems in distributed systems and to determine optimal agent-based solutions.  We showed that from a cost-optimal strategy the system components that actually need to be purchased can be derived.  Our technique is based on a propositional logic encoding of the synthesis problem.  A truth assignment that maximises the sum of weights of satisfied clauses of the encoding characterises an optimal strategy.  This enables us to exploit the power of state-of-the-art Max-SAT tools to solve the synthesis problem. 
We showed that optimal winning strategies synthesised via Max-SAT can involve significant cost savings in comparison to bare winning strategies synthesised via standard SAT.

As future work we plan to consider enhanced goal properties allowing for goal-dependencies (with regard to the order in which goals have to be achieved) and for periodic goals. For the latter, we intend to combine our synthesis technique with $k$-induction \cite{timm2020model} in order to handle strategies that result in a looped run of the system.  While our MRAs are an abstract concept for representing resource allocation problems,  they can be easily adjusted to model concrete problems in distributed computing.  Thus, we plan to apply our approach to real-world scenarios in the fields of distributed operating systems \cite{kurose1989microeconomic},  wireless sensor networks \cite{mukherjee2019adai} or clouds \cite{chang2010optimal}.  
A further direction of future work is to tune the Max-SAT solver by employing heuristics that exploit the particular structure of our encodings of cost-optimal strategy synthesis problems.

\bibliographystyle{eptcs}
\bibliography{references}
\end{document}